\documentclass[twocolumn,showpacs,floatfix,aps]{revtex4}
\usepackage[dvips]{graphics}

\def \beq {\begin{equation}}
\def \eeq {\end{equation}}

\usepackage{amssymb}

\begin{document}

\title{Accretion of nonminimally coupled scalar fields into   black holes}

\author{Manuela G. Rodrigues}
\email{manu@ifi.unicamp.br}
\affiliation{ Instituto de F\'\i sica Gleb Wataghin, UNICAMP,
C.P. 6165, 13083-970 Campinas, SP, Brazil.}
\author{Alberto Saa}
\email{asaa@ime.unicamp.br}
\affiliation{
Departamento de Matem\'atica Aplicada,
  UNICAMP,
C.P. 6065, 13083-859 Campinas, SP, Brazil.}

\begin{abstract}
By using a quasi-stationary approach,
we consider the mass evolution of Schwarzschild black holes in the presence of
a nonminimally coupled cosmological
scalar field.
The mass evolution equation is analytically solved for   generic
  coupling, revealing a
qualitatively distinct behavior from the  minimal coupling case. In particular,
for black hole masses smaller than a certain critical value,
the accretion of the   scalar field can lead to
mass decreasing
 even if no phantom energy is involved.
 The physical validity
of the adopted  quasi-stationary approach and some implications of our result
for the evolution of primordial and astrophysical black holes
are discussed. More precisely, we argue that black hole  observational
data could be used to place constraints on the nonminimally  coupled
energy content of the universe.
\end{abstract}

\pacs{ 04.70.Bw, 95.36.+x, 97.60.Lf,  98.80.Cq}
\maketitle

\section{Introduction}

The accretion of matter is one of the most studied
physical process involving black holes. Assuming the validity
of certain energy
conditions for the accreting matter, the  black  hole mass will never decrease.
In fact, if the null energy
condition holds, no classical process can lead to mass decreasing for black
holes\cite{HawkingEllis}.
The situation changes completely
 if quantum processes are allowed:
a black hole can, in fact, shrink
due to the emission of
 Hawking radiation\cite{HawkingRadiation}.
 Such processes are particularly relevant, for instance, to
Primordial black holes (PBH)\cite{CarrReview}.
One of the most striking features of PBH is that
they could indeed
evaporate completely due to the emission of Hawking radiation. It is known,
in particular, that
a PBH with mass smaller than the so called
Hawking mass $M_{\rm H} = 10^{15}$g  should have already evaporated
 by now. PBH with masses close to that limit are specially relevant
 because their emitted Hawking radiation might, in principle,
produce observable effects in the present day universe\cite{PBHsearch}.

The interest in these   problems  has increased considerably
in the last years due to  the many  dark energy phenomenological
models that have been proposed to described
the recent accelerated expansion of the universe\cite{accel}.
Such
models\cite{Quintessence} typically involve a scalar field pervading
all the universe that could, in principle, be absorbed
by any black hole, implying consequently
in new channels for black hole mass accretion\cite{Magueijo}.
It is interesting to notice that
the study of   black holes growth  in the presence of scalar fields
has been   initiated before\cite{BarrowCarr} the discovery of the recent
acceleration of the universe and, thus,  before the proposal of any dark energy
model.

The mass evolution of any black hole is   governed   by two competing
 processes.  The first one is Hawking radiation, which decreases
 the black hole mass due to the emission of a thermal
 radiation. The other one, which tends to increase the black hole
 mass, is the accretion of the surrounding available matter and
 energy.  The survival or not of a PBH until nowadays, for instance,
  will depend
 on the detailed  balance of these processes.
  The unexpected possibility that
   black hole masses could effectively
   decrease due to the accretion of exotic
 (phantom) dark energy\cite{Babichev} was received with great
 interest because, mainly, it could alter qualitatively the evolution of
 any black hole, implying, occasionally, in observational consequences
  for both astrophysical and primordial black holes.
Since phantom dark energy violates the usual energy conditions,
there is no contradiction between these results and the classical
theory of black holes.
  Nevertheless, one should keep in mind that
  the physical viability of models involving phantom energy has
  been
 constantly challenged by their severe inherent classical
  and quantum instabilities\cite{instabilities}.

In this paper, we   study   the mass evolution
of Schwarzschild black holes in the presence of a nonminimally coupled
scalar field.
A quasi-stationary
approach is introduced and the
mass evolution equation is analytically solved for generic
coupling.
Our main conclusion  is   that, for black hole initial  masses smaller
 than a certain critical value, one could indeed have mass decreasing even
 in the absence of the Hawking evaporation mechanism and
   without   any component of phantom energy in the model.
This is a more robust scenario for mass decreasing of black holes
due to the accretion of exotic matter since it is not plagued by the
phantom energy instabilities. Moreover, one could have, in principle,
 mass decreasing
for considerably larger black holes than the minimally coupled case,
with possible implications for primordial and astrophysical black holes,
which could be explored in order to place observation constraints
on  the nonminimally coupled energy content of the universe.

\section{Nonminimally coupled scalar fields around   black holes}

 We are  concerned  here with a scalar field $\phi$  governed by the action
\beq
\label{nmc}
S = \frac{1}{2}\int d^4x\sqrt{-g}\left[F(\phi)R - \partial_a\phi\partial^a\phi - 2V(\phi) \right],
\eeq
surrounding a Schwarzschild black hole. Nonminimally coupled
cosmological models of the type (\ref{nmc}) have been
intensively used in modern cosmology\cite{nmc}.
Models for which it is indeed possible to reach $F(\phi)=0$ are known
to be plagued with singularities\cite{sing}. The hypersurface
$F(\phi)=0$ marks, in a sense, the boundary between standard
($F(\phi)>0$) and phantom-like ($F(\phi)<0$) behavior for the
scalar field $\phi$\cite{abramo}. We are mainly interested here in models such
that $F(\phi)>0$ everywhere since, in such cases, phantom-like behavior
is excluded by construction.

 Since   Schwarzschild spacetime is Ricci-flat,
 the equation of motion
for $\phi$ obtained from (\ref{nmc}) reads simply
\beq
\label{KG}
\Box \phi  = V'(\phi),
\eeq
and the associated energy momentum tensor is given by
\beq
T_{ab} = \partial_a\phi\partial_b\phi  - \frac{g_{ab}}{2} \left(
\partial_c\phi\partial^c\phi +2 V \right) +  \nabla_a  \nabla_b F - g_{ab}\Box F.
\eeq
Note that, due to Ricci-flatness of Schwarzschild spacetime, we
have $\nabla_bT_{a}^{\phantom{a}b}=0$.
By adopting the usual Schwarzschild coordinates, the spherically
 symmetrical version of Eq. (\ref{KG}) will
be given by
\begin{eqnarray}
\label{KG1}
    -\frac{\partial^2\phi}{\partial t^2} +
\frac{1}{r^2}\left(1-\frac{2M}{r}   \right) \frac{\partial}{\partial r}  \left[
 r^2\left(1-\frac{2M}{r}    \right) \frac{\partial\phi}{\partial r} \right]
 &=& \nonumber \\ \left( 1-\frac{2M}{r}    \right) V'(\phi). &&
\end{eqnarray}
The standard formulation of the stationary
Bondi accretion process\cite{BondiAccretion} for this problem
consists in considering    solutions of (\ref{KG1}) with the following
boundary condition
\beq
\label{bc}
\lim_{r\rightarrow\infty }\phi(t,r) = \phi_{c}(t),
\eeq
where $\phi_{c}(t)$ corresponds to the cosmological homogeneous
and isotropic solution of the model (\ref{nmc}), with   cosmological  and
Schwarzschild time coordinates identified. Since no back reaction of the
scalar field is taken into account, our approach requires that
 the energy content of the scalar
field must remain bounded and small around the black hole.
Once we have a   solution $\phi(t,r)$  of (\ref{KG1}) with bounded energy
and obeying the boundary
condition (\ref{bc}), we assume that its energy flux on the
black hole horizon   is completely absorbed by the black
hole, implying that
\beq
\frac{dM}{dt} = \oint_{r=2M} r^2 T_{t}^{\phantom{t}r}  d\Omega\,    .
\eeq
 This problem was solved, for $F(\phi)=1$ and $V(\phi)=0$, in \cite{Jacobson}.
In the Eddington-Finkelstein coordinates $(v, r)$, with
$v = t + r + 2M\log\left(r/2M -1 \right)$  corresponding to
incoming light geodesics,  the pertinent solution    corresponds
to the stationary configuration
\beq
\label{stationary}
 \phi(v,r)=\beta + \gamma\left(  v-r+2M\log \frac{2M}{r}\right) ,
 \eeq
with $\beta$ and $\gamma$ constant. We do not expect to have
 stationary solutions like this for the generic model  (\ref{nmc}).
 In fact,   stationary solutions are possible only for actions that are
 invariant under shifts $\phi\rightarrow \phi + \lambda$, see \cite{Stationary}.
 We can, however, adopt a quasi-stationary approach based on the
 observation\cite{FrolovKofman} that, for slowly varying cosmological
 solutions $\phi_c(t)$, the ``delayed'' field configuration given by
  \beq
  \label{sol}
\phi(v,r) = \phi_c\left( v-r+2M\log\frac{2M}{r}\right),
\eeq
is an approximated
solution of (\ref{KG1}) for certain potentials $V(\phi)$. The validity   of this
approximation will assure, of course,  the validity of our quasi-stationary
approach. By substituting  (\ref{sol})  in (\ref{KG1})  one
gets
\beq
\label{cond}
\left( 1 +  \frac{2M}{r}  + \left( \frac{2M}{r} \right)^2 + \left( \frac{2M}{r} \right)^3 \right) \ddot\phi_c +  V'(\phi_c) = 0,
\eeq
with the dot standing for the derivative with respect to $t$. Hence,
our approximation is  valid if $\ddot\phi_c\approx 0$ and
$V'(\phi_c) \approx 0$.
Due to the typical cosmological time scales, the assumption of
a quasi-stationary  ($\ddot\phi_c\approx 0$) evolution around
the black hole is not, in fact,
too restrictive.
The same is true for the assumption
 $V'(\phi_c) \approx 0$, but the argument is more
involved. Assuming a small variation of $\phi_c$, the
potential can be linearized as $V(\phi_c) =   \mu \phi_c$, since
the constant factor is irrelevant here. In this case, equation
(\ref{KG1}) will be   a linear equation, and it is possible
to find a stationary solution   obeying
the Bondi boundary condition (\ref{bc}).
The approximation will be valid provided
$\phi_c$ is small and $r$ is kept smaller than the cosmological
horizon scale,   see \cite{Magueijo}
for the details.
It is interesting to notice that the explicit examples of failure
of the approximation (\ref{sol})  presented
in \cite{FrolovKofman} corresponds clearly to situations where
one cannot assure  $\ddot\phi_c  \approx 0$ or $V'(\phi_c) \approx 0$.

For the solution (\ref{sol}), one has
\beq
\label{emt}
T_{t}^{\, r} = \left(\frac{2M}{r}\right)^2
\left( \left(1+F'' \right)\dot\phi_c^2  +
F'\ddot\phi_c- \frac{F'}{4M}\dot\phi_c \right).
\eeq
Also from (\ref{sol}), we see that,
on the black hole horizon, the field $\phi$ assumes the value of $\phi_c$,
propagated along a incoming light geodesic, but arriving with a
certain
``delay''\cite{FrolovKofman}. Our quasi-stationary
analysis neglects also such delay and,
hence, in the quasi-stationary approximation
\beq
\phi_c(t) \approx \phi_\infty + \dot{\phi}_\infty(t-t_0),
\eeq
with $\phi_\infty$ and $\dot{\phi}_\infty$ constants,
 we have
\beq
\label{accret}
\dot M = 16\pi M^2  \left(1+F'' \right)\dot\phi_\infty^2 -
4\pi M F' \dot\phi_\infty.
\eeq
For the minimal coupling case, $F(\phi)=1$ and (\ref{accret}) reduces to the
usual scalar field accretion rate\cite{Jacobson}. It is clear, however,
that for the nonminimally coupled case one could have, in principle,
$\dot M < 0$ even in the absence of phantom modes.
The rate (\ref{accret}) corresponds only to the
accretion of the scalar field.
The complete mass evolution equation is
obtained by adding to the right-handed side a term $\propto M^{-2}$
corresponding to the Hawking radiation.
As we will see in the next section,
the fact that the two accretion terms in (\ref{accret})  have different signs
and different powers of $M$
will imply in the existence of a critical mass $M_{\rm cr}$ delimiting
the mass increasing and decreasing accretion regimes.

We finish this section by noticing
 that the possibility of negative energy fluxes for
nonminimally coupled scalar fields and their implications for
mass decreasing process involving
black holes has been already considered
previously in another context, namely in the investigation
 of the generalized second law of thermodynamics\cite{Ford}.

\section{Mass evolution}

For a generic coupling function $F(\phi)$, the complete mass evolution equation
 has the general
form
\beq
\label{accret1}
\dot M = f(t) M^2  - g(t)M  - \frac{\alpha}{M^2},
\eeq
where $f(t)$ and $g(t)$ are smooth functions and
  $\alpha$ is a characteristic constant for  Hawking radiation.
Let us consider, initially, only the accretion process $(\alpha=0)$.
 By introducing
$M(t) = G(t)P(t)$, with
\beq
\label{G}
G(t) = e^{-\int_{t_0}^t g(s)\,ds},
\eeq
we obtain a separable equation for $P(t)$,  which can be easily solved
leading to the following solution for (\ref{accret1}) with $\alpha=0$
\beq
\label{solg}
M(t) = \frac{M_0G(t)}{1 - M_0H(t)},
\eeq
where $M(t_0)=M_0$ and
\beq
H(t) = \int_{t_0}^tf(s)G(s)\, ds.
\eeq
 Typically, if
the denominator of (\ref{solg}) does not vanish,
the mass $M(t)$ decreases   according to (\ref{G}) for positive
$g(t)$. Mass
increasing solutions appear  when the denominator vanishes.
For positive and well behaved  $f(t)$ and $g(t)$, the
function $H(t)$ will be monotonically increasing and bounded by
$ H_\infty = \lim_{t\to\infty} H(t) $, leading to
a critical mass $M_{\rm cr}  =  H_\infty^{-1}$.
Any black hole with   initial mass $M_0$ such that $0 < M_0 < M_{\rm cr}$,
even in the
absence of Hawking radiation,
will disappear due to the accretion of the scalar field,
but such process typically will take an infinite amount of time.
 On the other
hand, those black holes with  initial masses $M_0>M_{\rm cr}$ will grow by
accreting the scalar field. In fact, in this case,
the denominator of (\ref{solg}) vanishes for $t =   t_{\rm cr}$, with
$H(t_{\rm cr}) =  {M_0}^{-1},$
implying that the black hole grows up to infinite mass in a
finite time. The larger is the black hole initial mass $M_0$,
the shorter is $t_{\rm cr}$. In contrast to the $0< M_0 < M_{\rm cr}$
case, such behavior for  $M_0 > M_{\rm cr}$ is similar to that one observed
for the minimally coupled case $F=1$. The qualitative evolution for the case
 $M_0 = M_{\rm cr}$ will
depend on the details of the functions $f(t)$ and $g(t)$.

For situations with large $M_{\rm cr}$, the inclusion of
 Hawking radiation will alter
 qualitatively only the final instants
of the mass decreasing process. In such a case,
for $  M_0  < M_{\rm cr} $,
the black hole also disappears, but now in a finite time, since Hawking
radiation dominates the process  for  $M(t)\ll 1$. In fact,
for $M> M_{\rm cr}$, the Hawking radiation term can be
 neglected and the dynamics are essentially that one described
 by (\ref{solg}).
Let us now consider
some explicit examples of the coupling function $F(\phi)$
in order to elucidate these
points.

\subsection{$F(\phi) = 1 +\xi\phi$}

 In this linear coupling
case,
equation (\ref{accret1}) is autonomous, with
$f(t) = 16\pi \dot\phi_\infty^2$ and $g(t)=4\pi \xi  \dot\phi_\infty$,
and   can be integrated by quadrature  for any
value of $\alpha$.  We do
not need, however, the exact solution here.
We assume  $\xi$ and $\phi$  to be
both positive in order
to avoid   possible singularities\cite{sing} and, without loss of
generality, $t_0=0$.
The functions $G(t)$ and $H(t)$ are in
 this case
\beq
\label{g(t)}
G(t) =  e^{-4\pi \xi  \dot\phi_\infty t}
\eeq
and
\beq
H(t) =
\frac{4\dot\phi_\infty}{\xi}\left(1 - G(t)  \right).
\eeq
For   $\dot{\phi}_\infty$   positive,
we have
\beq
M_{\rm cr} =  \frac{\xi}{4} \dot{\phi}_\infty^{-1},
\eeq
 and
\beq
   t_{\rm cr} =  \frac{1}{4\pi \xi  \dot\phi_\infty}\log \frac{M_0}{M_0-M_{\rm cr}}.
\eeq
Notice that, for typical cosmological situations, $\dot\phi_\infty$ is small, implying in  large values of $M_{\rm cr}$ for $\xi$ of the   order of
unity  (in Planck
units). In these cases, the Hawking radiation is important only in
the final instants of the mass decreasing phase.

\subsection{$F(\phi) = 1 +\xi\phi^2$}

We assume $\xi >0$. We have $f(t)=16\pi(1+2\xi)\dot\phi_\infty^2$ and $g(t)=8\pi\xi\left(
\phi_\infty\dot\phi_\infty  + \dot\phi_\infty^2 t \right)$ in this case.
The pertinent  functions   are, for $t_0=0$,
\beq
G(t) = e^{-4\pi\xi\left(2\phi_\infty\dot\phi_\infty t
+  \dot\phi_\infty^2t^2 \right)}
\eeq
and
\beq
H(t) = 16\pi(1+2\xi)\dot\phi_\infty^2
\int_0^t  e^{-4\pi\xi\left(2\phi_\infty\dot\phi_\infty s
+  \dot\phi_\infty^2s^2 \right) } \, ds.
\eeq
The critical mass is given by $M_{\rm cr} =  H_\infty^{-1}$,
with
\beq
H_\infty = 4 {\pi} \frac{1+2\xi}{ \sqrt{\xi}}\left|\dot\phi_\infty\right| e^{4\pi\xi\phi_\infty^2}\left[ 1 - \sigma{\rm erf}\left( 2\sqrt{\pi\xi}\phi_\infty \right) \right] ,
\eeq
where $\sigma = {\rm sgn} \dot{\phi}_\infty$ and erf$(x)$ is the error function\cite{Gradshteyn}.
For the typical cosmological situations we have that    $ \phi_\infty$
is very small, leading   to
\beq
M_{\rm cr}\approx \frac{\sqrt{ \xi}}{4\pi(1+2\xi)}\left|\dot\phi_\infty\right|^{-1}.
\eeq
Notice that, as in the previous case, $M_{\rm cr}\propto \dot\phi_\infty^{-1}.$

\subsection{$F(\phi) = e^{\xi\phi}$}

In this case,  we have
$f(t)=16\pi\left(\! 1\!+\xi^2 e^{\xi(\phi_\infty   + \dot\phi_\infty t )}\right)\!\dot\phi_\infty^2$
 and $g(t)=4\pi\xi \dot\phi_\infty e^{\xi(\phi_\infty   + \dot\phi_\infty t )}$,
  leading, for $t_0=0$, to
\beq
G(t) = \exp\left(-4\pi e^{\xi\phi_\infty}\left(e^{\xi\dot\phi_\infty t} -1\right)\right)
\eeq
and
\beq
H(t) = 16\pi \dot\phi_\infty^2
\int_0^t \left(1+\xi^2e^{\xi(\phi_\infty
+  \dot\phi_\infty s  )} \right) G(s) ds.
\eeq
The critical mass is given by
\beq
M_{\rm cr}^{-1}  = \frac{16\pi \dot\phi_\infty}{\xi} \left[\frac{\xi^2}{4\pi} + \exp\left(
4\pi e^{\xi\phi_\infty}\right)
\Gamma\left(0,4\pi e^{\xi\phi_\infty} \right)\right],
\eeq
where $\Gamma(z,x)$ is the incomplete Gamma function\cite{Gradshteyn}.
For
$ \phi_\infty$
  small, we have
\beq
M_{\rm cr}\approx \frac{ {\xi}}{ a+4\xi^2 }\dot\phi_\infty^{-1},
\eeq
where $a$ is a numerical constant of the order of unity, namely
$a =
 16\pi e^{4\pi}\Gamma(0,4\pi) \approx
3.72$. Again, we observe  the same behavior $M_{\rm cr}\propto \dot\phi_\infty^{-1}.$

\subsection{Radiation era with $F(\phi) = 1 +\xi\phi$}

The previous examples involve only the nonminimally scalar field
in the quasi-stationary approximation. This is not enough, for
instance, to describe  PBH, since they were created in the
primordial universe and have existed for eras where dark energy
was not the gravitationally dominant content of the universe. In the radiation dominated era,
in particular, the universe was filled and dominated by   ultra relativistic
matter which energy density is described in Planck units by
\beq
\label{egamma}
\varepsilon_\gamma = \frac{3}{32\pi t^2}.
\eeq
Such an energy density has been also available to be accreted by
the black hole and should be incorporated in our analysis.
The case of linear coupling $F(\phi) = 1 +\xi\phi$ in the
presence of radiation with energy density (\ref{egamma})
 corresponds to the choices
$f(t) =
16\pi      \dot\phi_\infty^2 +  (3/2) t^{-2}  $ and $g(t)= 4\pi   \xi \dot\phi_\infty$. The $G(t)$ and $H(t)$ functions in this case
are
\beq
G(t) =  e^{-4\pi \xi  \dot\phi_\infty ( t - t_0)}
\eeq
and
\beq
H(t) = \frac{4\dot\phi_\infty}{\xi}\left(1 - G(t)  \right) +  \frac{3}{2} \int_{t_0}^t   s^{-2}     e^{-4\pi \xi  \dot\phi_\infty ( s - t_0)} \,  ds  ,
\eeq
leading to
\beq
\label{beta}
H_\infty =   \frac{4\dot\phi_\infty}{\xi}
\left( 1 + \frac{3 \xi e^\beta }{8\dot\phi_\infty t_0}  \beta   \Gamma(-1,\beta)  \right),
\eeq
with $\beta= 4\pi\xi\dot\phi_\infty t_0$. Since
\beq
\lim_{x\to 0} x\Gamma(-1,x) = 1,
\eeq
we have in the present  case
\beq
M_{\rm cr} = H_\infty^{-1} \approx
\frac{\xi}{4\dot\phi_\infty}\left(1 +
\frac{3 \xi}{8\dot\phi_\infty t_0}\right)^{-1} ,
\eeq
if $\beta$ is small.

\section{Discussion}

If we assume that $\dot\phi_\infty^2$ is of the same order of
the critical density of the universe today
 ($\rho_0 \approx 10^{-29}$g/cm$^3$), we have $M_{\rm cr}\approx
10^{56} g$ for coupling constants $\xi$ of the order of unity   (in Planck
units) in the three first
  cases considered in  the last section, allowing all the black
holes in the universe to be in the shrinking phase today.
In fact, even  the galactic   supermassive black holes (SMBH) with
$M\approx 10^6M_\odot \approx 10^{39}$g are far below such a limit.
These black holes would be  shrinking today  according to (\ref{solg}). The exact
characteristic decaying time will depend on the particular
coupling function. For the case of the linear coupling, the
characteristic time is, according to (\ref{g(t)}), $10^{17}$s, similar
to the universe age. Notice that all the other coupling functions considered
in the last section lead, typically, to faster decreasing mass regimes.

The fact
that there    are likely many black hole around us might  be used to
constraint the nonminimally coupled energy content of the universe during
the cosmological history.
Let us consider, for simplicity, the last example of the previous section:
the linear coupling case during the radiation dominated
era.   Suppose that the dark energy content of the universe has changed lightly after, say, $t_0 = 1$s.
In this case, $\dot\phi_\infty t_0 \approx 10^{-18}$ in Planck units,
justifying to take $\beta\approx 0$ in (\ref{beta}) and leading to
$M_{\rm cr}    \approx 10^{38}{\rm g}
$
for a coupling constant $\xi$ of the order of unity.
 Thus, only PBH with
mass greater than $10^{38}$g would escape from the shrink phase.
 Notice that this mass is extremely  large if compared with  the usual Hawking mass $M_{\rm H} = 10^{15}$g. Observational
constraints on the PBH mass cutoff\cite{PBHsearch} could be used, in principle,
 to establish constraints
on the non-minimal coupling parameter $\xi$, although the details
depends on the coupling function $F(\phi)$.
If we take $t_0 = 10^{11}$s, corresponding to the radiation-matter equality era,
we will have $\dot\phi_\infty t_0 \approx 10^{-7}$, leading to
$M_{\rm cr}    \approx 10^{49}{\rm g}$. This is, again, a huge mass and
implies
 that virtually all black holes
present at the end of radiation era have existed during all the matter dominated
era in a shrinking regime.  They should have lost two thirds of their
mass by now, suggesting that
observational data about SMBH could also be used to constraint
the nonminimally coupled energy content of the universe.

We finish by
  noticing two points. First, one knows that it is not
  expected, in general,  to have constant values for $\phi_\infty$ and $\dot\phi_\infty$
along the cosmological history. Equation (\ref{accret1})   accommodates also
situations where $\phi_\infty$ and $\dot\phi_\infty$ are functions of $t$.
However, we should    keep in mind that our formalism is based on the
assumption of
a quasi-stationary   evolution, requiring
$\ddot\phi_c(t)\approx 0$ in order to work properly.
One needs to take backreaction into account in order to treat
non stationary situations, see, for instance, \cite{guzman}  for
a recent discussion.

The second point is related with the hypothesis that $\phi$ is a field
test around a Schwarzschild black-hole. This is a good approximation
provided that  the energy content of the scalar field (dark energy) is
negligible when compared with the black-hole Physics scale. For the much
larger cosmological scale, on the other hand,
 the scalar field is indeed the dominant
energy content, being the sole responsible for the accelerated expansion
of the universe, usually
 described by a quasi-de Sitter solution. In our universe,
these two scales are very different. Since the dark energy content
is so small, in order to probe the quasi-de Sitter properties of the
spacetime one needs to consider length scales of the same order of the
Hubble radius. It is perfectly  possible, in particular, to apply condition (\ref{cond})
in a region far from de black-hole (large $r$), but still far from the
cosmological horizon.
Furthermore, provided that the effective cosmological constant of the
accelerated expansion is small, the dynamics near the black-hole
horizon are essentially the same of the Schwarzschild case, implying
that (\ref{accret}) is still valid.
 From a theoretical point of view, however,
it is certainly interesting to consider the problem of accretion
onto Schwarzschild-de Sitter black-holes as it is done, for instance,
in \cite{MartinMoruno} for the case of perfect fluids
and minimally coupled fields. We already know, however,
 that our present analysis
should arise naturally in the limit of small $\Lambda$.
These points are now under investigation.

\acknowledgements

This work was supported by FAPESP and CNPq.


\begin{references}

\bibitem{HawkingEllis} S.W. Hawking and G.F.R. Ellis, {\em The Large Scale Structure
of Space Time}, Cambridge University Press (1973).


 \bibitem{HawkingRadiation} S. W. Hawking, Nature {\bf 248}, 30 (1974);
 Comm. Mat. Phys. {\bf 43}, 199 (1975).

\bibitem{CarrReview}B.J. Carr, {\em
Primordial Black Holes as a Probe of Cosmology and High Energy Physics},
in {\em Quantum Gravity: From Theory to Experimental Search},
Ed. D. Giulini, C. Kiefer, and C. Lammerzahl, Lect. Notes Phys. {\bf 631},
 301 (2003),  [arXiv:astro-ph/0310838].


 \bibitem{PBHsearch} M. Schroedter, {\em et al.}, Astropart. Phys.
 {\bf 31}, 102 (2009).



\bibitem{accel} A. G. Riess, {\em et al.}, Astron. J. {\bf 116}, 1009 (1998);
S. Perlmutter, {\em
et al.}, Astrophys. J. {\bf 517}, 565 (1999).

\bibitem{Quintessence}  P. J. E. Peebles and B. Ratra,
Rev. Mod. Phys. {\bf 75}, 559
(2003); T. Padmanabhan, Phys. Rept. {\bf 380}, 235 (2003).

\bibitem{Magueijo}  R.~Bean and J.~Magueijo,
  %``Could supermassive black holes be quintessential primordial black
  %holes?,''
  Phys.\ Rev.\  D {\bf 66}, 063505 (2002).
  %%CITATION = PHRVA,D66,063505;%%

\bibitem{BarrowCarr} J.D. Barrow, Phys. Rev. D {\bf 46}, R3227 (1992);
J.D. Barrow and B.J. Carr, Phys. Rev. D {\bf 54}, 3920 (1996).

\bibitem{Babichev}
  E.~Babichev, V.~Dokuchaev and Yu.~Eroshenko,
  %``Black hole mass decreasing due to phantom energy accretion,''
  Phys.\ Rev.\ Lett.\  {\bf 93}, 021102 (2004);
  %%CITATION = PRLTA,93,021102;%%
  J.\ Exp.\ Theor.\ Phys.\  {\bf 100}, 528 (2005)
  [Zh.\ Eksp.\ Teor.\ Fiz.\  {\bf 127}, 597 (2005)].
  %%CITATION = ZETFA,127,597;%%

\bibitem{instabilities}  S.~M.~Carroll, M.~Hoffman and M.~Trodden,
  %``Can the dark energy equation-of-state parameter w be less than -1?,''
  Phys.\ Rev.\  D {\bf 68}, 023509 (2003);
  %%CITATION = PHRVA,D68,023509;%%
  S.~D.~H.~Hsu, A.~Jenkins and M.~B.~Wise,
  %``Gradient instability for w<-1,''
  Phys.\ Lett.\  B {\bf 597}, 270 (2004).
  %%CITATION = PHLTA,B597,270;%%

\bibitem{nmc}  T. Futamase and K.I. Maeda, Phys. Rev. {\bf D39},
399 (1989);
  %%CITATION = PHRVA,D39,399;%%
T. Futamase, T. Rothman, and R. Matzner,
 Phys. Rev. {\bf D39}, 405 (1989);
%%CITATION = PHRVA,D39,405;%%
S. Deser, Phys. Lett. {\bf 134B}, 419 (1984);
%%CITATION = PHLTA,B134,419;%%
Y. Hosotani, Phys. Rev. {\bf D32}, 1949 (1985);
%%CITATION = PHRVA,D32,1949;%%
O. Bertolami, Phys. Lett. {\bf 186B}, 161 (1987);
%%CITATION = PHLTA,B186,161;%%
  S.~Sonego and V.~Faraoni,
  %``Coupling to the curvature for a scalar field from the equivalence
  %principle,''
  Class.\ Quant.\ Grav.\  {\bf 10}, 1185 (1993);
  %%CITATION = CQGRD,10,1185;%%
  V.~Faraoni,
  %``Non-minimal coupling of the scalar field and inflation,''
  Phys.\ Rev.\  D {\bf 53}, 6813 (1996);
  %%CITATION = PHRVA,D53,6813;%%
    N.~Bartolo and M.~Pietroni,
  %``Scalar-Tensor Gravity and Quintessence,''
  Phys.\ Rev.\  D {\bf 61}, 023518 (1999).
  %%CITATION = PHRVA,D61,023518;%%
  C.~Baccigalupi, S.~Matarrese and F.~Perrotta,
  %``Tracking extended quintessence,''
  Phys.\ Rev.\  D {\bf 62}, 123510 (2000);
  %%CITATION = PHRVA,D62,123510;%%
  E.~Gunzig, A.~Saa, L.~Brenig, V.~Faraoni, T.~M.~Rocha Filho and A.~Figueiredo,
  %``Superinflation, quintessence, and nonsingular cosmologies,''
  Phys.\ Rev.\  D {\bf 63}, 067301 (2001);
  %%CITATION = PHRVA,D63,067301;%%
  A.~Saa, E.~Gunzig, L.~Brenig, V.~Faraoni, T.~M.~Rocha Filho and A.~Figueiredo,
  %``Superinflation, quintessence, and the avoidance of the initial
  %singularity,''
  Int.\ J.\ Theor.\ Phys.\  {\bf 40}, 2295 (2001);
  %%CITATION = IJTPB,40,2295;%%
  V.~Faraoni,
  %``A crucial ingredient of inflation,''
  Int.\ J.\ Theor.\ Phys.\  {\bf 40}, 2259 (2001);
  %%CITATION = IJTPB,40,2259;%%
F.~C.~Carvalho and A.~Saa,
  %``Non-minimal coupling, exponential potentials and the w < -1 regime of  dark
  %energy,''
  Phys.\ Rev.\  D {\bf 70}, 087302 (2004).
  %%CITATION = PHRVA,D70,087302;%%

\bibitem{sing}
  L.~R.~Abramo, L.~Brenig, E.~Gunzig and A.~Saa,
  %``On the singularities of gravity in the presence of non-minimally coupled
  %scalar fields,''
  Phys.\ Rev.\  D {\bf 67}, 027301 (2003);
  %%CITATION = PHRVA,D67,027301;%%
  L.~R.~Abramo, L.~Brenig, E.~Gunzig and A.~Saa,
  %``Dynamical study of the singularities of gravity in the presence of
  %non-minimally coupled scalar fields,''
  Int.\ J.\ Theor.\ Phys.\  {\bf 42}, 1145 (2003);
  %%CITATION = IJTPB,42,1145;%%
L.~A.~Elias and A.~Saa,
  %``Homogeneous cosmologies and the Maupertuis-Jacobi principle,''
  Phys.\ Rev.\  D {\bf 75}, 107301 (2007);
  %%CITATION = PHRVA,D75,107301;%%
    M.~F.~Figueiro and A.~Saa,
  %{\em Anisotropic singularities in modified gravity models},
  Phys.\ Rev.\ D {\bf 80}, 063504 (2009).
  %%CITATION = ARXIV:0906.2588;%%

\bibitem{abramo}
  L.~R.~Abramo, L.~Brenig and E.~Gunzig,
  %``On the stability of gravity in the presence of a non-minimally coupled
  %scalar field,''
  Phys.\ Lett.\  B {\bf 549}, 13 (2002).
  %%CITATION = PHLTA,B549,13;%%


\bibitem{BondiAccretion}   H.~Bondi,
  %``On spherically symmetrical accretion,''
  Mon.\ Not.\ Roy.\ Astron.\ Soc.\  {\bf 112}, 195 (1952).
  %%CITATION = MNRAA,112,195;%%


\bibitem{Jacobson}
  T. Jacobson,
  %``Primordial black hole evolution in tensor scalar cosmology,''
  Phys.\ Rev.\ Lett.\  {\bf 83}, 2699 (1999).
  %%CITATION = PRLTA,83,2699;%%

\bibitem{Stationary} R. Akhoury, C.S. Gauthier, and A. Vikman,
JHEP {\bf 03}, 082 (2009).

\bibitem{FrolovKofman} A. Frolov and L. Kofman,
JCAP {\bf 0305}, 009  (2003).

\bibitem{Ford}
  L.~H.~Ford and T.~A.~Roman,
  %``Classical scalar fields and violations of the second law,''
  Phys.\ Rev.\  D {\bf 64}, 024023 (2001).
  %%CITATION = PHRVA,D64,024023;%%



\bibitem{Gradshteyn} I. S. Gradshteyn and
I. M. Ryzhik,
{\em Table of Integrals, Series, and Products},
Academic Press, (2007).

\bibitem{guzman}
  J.~A.~Gonzalez and F.~S.~Guzman,
  %``Accretion of phantom scalar field into a black hole,''
  Phys.\ Rev.\  D {\bf 79}, 121501R (2009).
  %%CITATION = PHRVA,D79,121501;%%

\bibitem{MartinMoruno}
  P.~Martin-Moruno, A.~E.~Marrakchi, S.~Robles-Perez and P.~F.~Gonzalez-Diaz,
  {\em Dark Energy Accretion onto black holes in a cosmic scenario},
 arXiv:0803.2005 [gr-qc]; to appear in Gen. Relat. Grav.
  %%CITATION = ARXIV:0803.2005;%%


\end{references}
\end{document}